\documentstyle[12pt]{article}

\begin{document}
\thispagestyle{empty}

\begin{center}
\LARGE \tt \bf{Riemannian and non-Riemannian geometries in filament rods and elastic walls}
\end{center}

\vspace{1cm}

\begin{center} {\large L.C. Garcia de Andrade\footnote{Departamento de
F\'{\i}sica Te\'{o}rica - Instituto de F\'{\i}sica - UERJ

Rua S\~{a}o Fco. Xavier 524, Rio de Janeiro, RJ

Maracan\~{a}, CEP:20550-003 , Brasil.

E-Mail.: garcia@dft.if.uerj.br}}
\end{center}

\vspace{1.0cm}

\begin{abstract}
The Riemannian geometry of elastica in one and two dimensions is considered. Two examples from engineering science are given. The first is the deflexion or Frenet curvature of the elastic filament rod where the Riemannian curvature vanishes, since the curve is one dimensional. However the Frenet curvature scalar appears on the Levi-Civita-Christoffel symbol of the Riemannian geometry. A second example is the bending of a planar two-dimensional wall where only the horizontal lines of the planar wall are bent, or a plastic deformation without cracks or fractures. In this case since the vertical lines are approximately not bent, and remain vertical while the horizontal lines are slightly bent in the limit of small deformations. This implies that the Gaussian  curvature vanishes. However the Riemann curvature does not vanish and again may be expressed in terms of the elastic properties of the planar wall. All computations were made possible using the GR-tensor package of the MAPLE V computer program. Non-Riemannian geometry in its own is applied to rods with nonhomogeneous cross-sections and computation of Cartan torsion in terms of the twist and the Riemann tensor are computed from the twist of the rod. In the homogeneous case the Riemann tensor maybe also obtained from Kirchhoff equations by comparing them to the non-geodesic equations and computing the affine connection. The external force acting on the rod represents the term which is responsible for the geodesic deviation in the space of the total Frenet curvature and twist. The Riemann tensor appears in terms of the mechanical torsional moment.   
\end{abstract}      
\vspace{1.0cm}       
\begin{center}
\Large{PACS number(s) : 0420,0450}
\end{center}

\newpage
\pagestyle{myheadings}
\markright{\underline{Riemannian geometry of elastica}}
\paragraph*{}
\section{Introduction}
Geometrical models in elasticity and plasticity have been proved useful in early years in engineering science, respectively given by Riemannian and non-Riemannian geometries respectively, mainly through the works of Kazuo Kondo, Bilby and K\"{o}ner \cite{1,2,3}. In  particular Bilby an metallurgical engineer of Shefield in 1955, made used of the non-Riemannian models to describe dislocations or defects in crystals. Bilby a planar model and Einstein's teleparallel theory where the Riemann curvature of the planar lattice. More recently Fonseca and Malta \cite{4,5} have considered the investigation of helical elastic filament rods where no fracture or cracks are developed. In their work they made use of the Kirchhoff equations which taking into account external forces. In this paper we consider two examples of Riemannian geometry of elastica in one and two dimensions without using the Kirchhoff equations and starting from simple bending of the rod according to Frenet curvature, where the metric is obtained. From the line element of the elastic bent rod we obtain the Levi-Civita affine connections which are symmetric and therefore does not present any Cartan torsion or Frenet torsion since the bending of the rod is planar and in principle does not present any twist. In this first Riemannian example the Riemann tensor vanishes since the model is one dimensional, while the Frenet scalar curvature appears explicitly in the one component Christoffel symbol. On a second example we also split a planar wall into orthogonal grid lines where only the horizontal lines are bent while the vertical lines appear almost intact in this plastic behaviour. The Riemann curvature tensor this time does not vanish and it is computed from the two dimensional metric of the elastica. Now the Riemann tensor is expressed in terms of the derivatives of the Frenet curvature of the planar elastic planar wall bent along the horizontal parallel directions. Non-Riemannian geometry of elastica \cite{6} is applied to Kirchhoff equations where the affine connection now is obtained from the non-geodesic equation where forces yield the geodesic deviation. Similar analog models such as the non-Riemannian vortex acoustics in fluids and superfluids have been recently obtained in the literature \cite{7}. Cartan torsion \cite{8} is also computed from the elastic properties of material as well as the Riemann components. Interesting from the mathematical point of view is that we do not, this time, obtain the metric straight from the Kirchhoff equations but solely the affine connection which is symmetric in the Riemann case. This procedure is similar to the one used by Kleinert \cite{9} in metric-affine non-Riemannian spaces to investigate the non-linear problem of spinning top. Actually as we shall see the first two examples are nonlinear since the approximation of slight deflexion of the rod is not assumed. The present paper is organised as follows: In section II we present the two Riemannian elaticity examples of deformation in one and two dimensions. In section III the non-Riemannian elastica is computed and the Kirchhoff equations are written in the form of a nongeodesic equations of motion where the coordinates are in fact the total Frenet curvatures along two diections while the third coordinate is the total twist along the filament curve. Section IV contains comments and future prospects.
\section{Riemannian elastica in  2-D and 3-D}
In this section we shall make use of the MAPLE V tensor package to compute the components of the Riemann tensor of one and two dimensional elastica from the metric obtained by the examples of deflexions of a filament rod and a planar elastic walls. Let us now consider the expression \cite{10} for the Frenet curvature of a filament rod slightly defleted by an angle ${\theta}$ given by
\begin{equation} 
\kappa= \frac{1}{\rho}= \frac{d{\theta}}{dx}
\label{1}
\end{equation}
where $dx$ is approximated equal to ds , the elementary arc lenght of the elastic filament. For angles ${\theta}<<1$ one obtains 
\begin{equation}
{\theta}=tan{\theta}=\frac{d{\nu}}{dx}={\nu}'
\label{2}
\end{equation}
where $\nu$ is the vertical deflexion and dash represents derivative with respecto x. Now the metric of the filament rod can be obtained by the expression
\begin{equation}
ds^{2}=dx^{2}+d{\nu}^{2}
\label{3}
\end{equation}
But since the deflexion $\nu$ does depemd on the variable x one can reduce the line element (\ref{3}) to the one dimensional line element of the rod
\begin{equation}
ds^{2}=[1+{{\nu}'}^{2}]dx^{2}
\label{4}
\end{equation}
Thus from this Riemannian line element one can obtain the metric component $g_{xx}=[1+{{\nu}'}^{2}]$. From this elastica metric component the Christoffel connection can be written as
\begin{equation}
{{\Gamma}^{x}}_{xx}=\frac{1}{2}g^{xx}g_{xx}= \kappa {\nu}'
\label{5}
\end{equation}
But note that there is a flaw in this computation since in this approximation ${{\nu}'}^{2}$ would vanish and therefore the metric would be one dimensional flat and the connection would vanish and obviously the Riemann curvature. The nonlinear case given by the planar elastic wall can be expressed by analogy by the metric 
\begin{equation}
ds^{2}=[1+{{\nu}'}^{2}]dx^{2}+dy^{2}
\label{6}
\end{equation}
where now the metric components are given by $g_{11}=[1+{{\nu}'}^{2}]$ and $g_{22}=1$. From these components we can use the MAPLE V tensor package to obtain the following expression by the Riemann tensor 
\begin{equation} 
R_{1212}= [{\partial}_{y}\kappa]^{2}+{\kappa}{\partial}_{y}\kappa
\label{7}
\end{equation}
where $\kappa$ and ${\nu}$ are respectively the Frenet curvature and deflexion along the y-direction. The decision to make just deflexion along the y-direction was to simplify computations since the complete elastica two dimensional metric would produce a huge expression for the Riemann tensor of exactly 16 terms. Other quantities such as Christoffel connections, and Ricci tensor and scalar can be easily computed by the tensor package. 
In the next section we shall see that problems found in the computations in this section can be erased by making use of the Kirchhoff equations of the rods. 
\section{Non-Riemannian elastic and Kirchhoff equations}
In this section we shall compute the Riemann tensor from the affine connection obtained by writting down the Kirchhoff equations for the elastic rods as nongeodesic equations. To establish this result let us remind the expressions for the Kirchhoff equations
\begin{equation}
{f_{1}}'-f_{2}k_{3}+f_{3}k_{2}=0
\label{8}
\end{equation}
\begin{equation}
{f_{2}}'+f_{1}k_{3}-f_{3}k_{1}=0
\label{9}
\end{equation}
\begin{equation}
{f_{3}}'-f_{1}k_{2}+f_{2}k_{1}=0
\label{10}
\end{equation}
\begin{equation}
{I(s)k_{1}}'+({\Gamma}-1)I(s)k_{2}k_{3}-f_{2}=0
\label{11}
\end{equation}
\begin{equation}
{I(s)k_{2}}'-({\Gamma}-1)I(s)k_{1}k_{3}+f_{1}=0
\label{12}
\end{equation}
\begin{equation}
(I(s)k_{3}{\Gamma})'=0
\label{13}
\end{equation}
where we $I(s)$ is the inertia moment, $\Gamma$ is the ratio between the Young's modulus and the shear modulus and the force is given in the director frame $(\vec{d}_{1},\vec{d}_{2},\vec{d}_{3})$ $(i=1,2,3)$, by $\vec{F}= f_{i}\vec{d}_{i}$ where the Einstein summation convention is implied and ${\kappa}_{F}=\sqrt{{k_{1}}^{2}+{k_{2}}^{2}}$ while $k_{3}$ is the twist density. It is easy to note that by writing the $k_{i}=\frac{dX_{i}}{ds}$ the Kirchhoff equations (\ref{11}) and (\ref{13}) can be expressed as the nongeodesic equations
\begin{equation}
\frac{d^{2}X_{1}}{ds^{2}}+({\Gamma}-1)\frac{dX_{2}}{ds}\frac{dX_{3}}{ds}-{I^{-1}(s)}\frac{dX_{2}}{ds}=-\frac{dlnI(s)}{ds}\frac{dX_{1}}{ds}
\label{14}
\end{equation}
Along with the analogous equation for $X_{2}$ we obtain the only non-vanishing Levi-Civita connection components as 
\begin{equation}
{{\Gamma}^{1}}_{23}=({\Gamma}-1)=-{{\Gamma}^{2}}_{13}
\label{15}
\end{equation}
From these Levi-Civita affine connections one obtains the only non-vanishing component of the Riemann tensor as
\begin{equation}
{R^{1}}_{313}=-(1-{\Gamma})^{2}
\label{16}
\end{equation}
Since the twist density $k_{3}$ can be expressed in terms of the torsional moment density $M_{3}$ as
\begin{equation}
k_{3}=\frac{dT}{ds}=\frac{M_{3}}{{\Gamma}I}
\label{17}
\end{equation}
where T is the total twist along s given by
\begin{equation}
T=\int{\frac{M_{3}}{{\Gamma}I(s)}ds}
\label{18}
\end{equation}
and ${\Gamma}$ is constant one may easily expressed the Riemann tensor component in terms of the total twist as
\begin{equation}
{R^{1}}_{313}=-(1-\frac{1}{T}\int{\frac{M_{3}}{I(s)}ds})^{2}
\label{19}
\end{equation}
Let us now consider the strain tensor of the nonhomogeneous rod the Kirchhoff theory as \cite{4}
\begin{equation}
{\epsilon}_{{\alpha}{\beta}}=\frac{1}{2}(\frac{{\partial}u_{\alpha}}{{\partial}X_{\beta}}+\frac{{\partial}u_{\beta}}{{\partial}X_{\alpha}})
\label{20}
\end{equation}
\begin{equation}
{\epsilon}_{{\alpha}3}=\frac{1}{2}(\frac{{\partial}u_{3}}{{\partial}X_{\alpha}}+(k_{3}-{k_{3}}^{0}X_{\alpha})
\label{21}
\end{equation}
\begin{equation}
{\epsilon}_{33}= (k_{1}-{k_{1}}^{0})X_{2}-(k_{2}-{k_{2}}^{0})X_{1}
\label{22}
\end{equation}
where $({\alpha}=1,2)$. From these expressions and the components of Cartan torsion \cite{10} 
\begin{equation}
S_{ijk}=\frac{1}{2}({\partial}_{i}{\partial}_{j}-{\partial}_{j}{\partial}_{i})u_{k}
\label{23}
\end{equation}
one may compute the Cartan torsion tensor components , which may be useful to investigate cracks and fractures on the rod
\begin{equation}
S_{{\beta}{\alpha}3}= [{\partial}_{\beta}{\epsilon}_{{\alpha}3}-{\partial}_{\alpha}{\epsilon}_{{\beta}3}]+[{\partial}_{\alpha}((k_{3}-{k_{3}}^{0})X_{\beta}-{\partial}_{\beta}((k_{3}-{k_{3}}^{0})X_{\alpha}
\label{24}
\end{equation}
\begin{equation}
S_{{\beta}{\alpha}3}= [{\partial}_{3}{\epsilon}_{j3}-{\partial}_{j}{\epsilon}_{33}-{{\partial}_{3}}^{2}u_{j}]
\label{25}
\end{equation}
Therefore we may conclude that the components of Cartan torsion may be expressed in terms of the twist density $k_{3}$. The expression for the elastic Riemann tensor of the nonhomogeneous rod can be expressed as 
\begin{equation}
{R_{ijkl}}=({\partial}_{i}{\partial}_{j}-{\partial}_{j}{\partial}_{i}){\partial}_{k}u_{l}
\label{26}
\end{equation}
and its components can be easily computed in terms of the twist density as 
\begin{equation}
{R_{ijkl}}=({\partial}_{i}{\partial}_{j}-{\partial}_{j}{\partial}_{i})[2{\epsilon}_{kl}-{\partial}_{l}u_{l}]
\label{27}
\end{equation}
\begin{equation}
{R_{ij{\alpha}3}}=({\partial}_{i}{\partial}_{j}-{\partial}_{j}{\partial}_{i})[(k_{3}-{k_{3}}^{0})X_{{\alpha}}]
\label{28}
\end{equation}
which reduces to
\begin{equation}
{R_{ij{\alpha}3}}=({\partial}_{i}{\partial}_{j}-{\partial}_{j}{\partial}_{i})[2{\epsilon}_{{\alpha}3}-{\partial}_{\alpha}u_{3}]
\label{29}
\end{equation}
while the affine elastic connection possesses the following symmetries
\begin{equation}
{{\Gamma}_{ijk}}=({\partial}_{i}{\partial}_{j})u_{k}
\label{30}
\end{equation}
\begin{equation}
{{\Gamma}_{ijk}}+{{\Gamma}_{ikj}}= 2({\partial}_{i}{\epsilon}_{jk})
\label{31}
\end{equation}
The geometrical models for the rods investigated in this section may be proved useful to investigate the fracture and cracks in  materials.

\section{Conclusions}
The ideas discussed and presented here maybe useful the invesitigation of the elastic problems in engineering and materials science as nanotubes for example. The recently investigation made by Ricca \cite{11} on the energy spectrum of the twisted flexible string under elastic relaxation such as torsional bending and bending energy may be in near future computed and built in terms of the geometrical models discussed here.

\section*{Acknowledgments}
\paragraph*{}
Thanks are due to C. Malta for helpful discussions on the subject of this paper. I am very much indebt to Universidade do Estado do Rio de Janeiro(UERJ) and CNPq. (Brazilian Government Agency) for financial support.

\end{document}